\documentclass[pre,aps,twocolumn,superscriptaddress,floatfix,amsmath,amssymb,showpacs]{revtex4}

\usepackage{graphicx}
\usepackage{bm}
\usepackage{hyperref}

\preprint{LA-UR 09-08205}

\begin{document}

\title{
  Statistical mechanics model for the transit free energy of monatomic liquids
}

\date{\today}

\author{Duane C. Wallace}
\author{Eric~D. Chisolm}
\author{N. Bock}
\author{G. De~Lorenzi-Venneri}
\affiliation{Theoretical Division, Los Alamos National Laboratory, Los Alamos,
NM 87545, USA}

\begin{abstract}

In applying vibration-transit (V-T) theory of liquid dynamics to the thermodynamic properties of monatomic liquids, the point has been reached where an improved model is needed for the small ($\sim\!10\%$) transit contribution.  Toward this goal, an analysis of the available high-temperature experimental entropy data for elemental liquids was recently completed [D.~C.~Wallace, E.~D.~Chisolm, and N.~Bock, Phys.\ Rev.\ E {\bf 79}, 051201 (2009)].  This analysis yields a common curve of transit entropy vs.~$T/\theta_{\rm tr}$, where $T$ is temperature and $\theta_{\rm tr}$ is a scaling temperature for each element.  In the present paper, a statistical mechanics model is constructed for the transit partition function, and is calibrated to the experimental transit entropy curve.  The model has two scalar parameters, and captures the temperature scaling of experiment.  The calibrated model fits the experimental liquid entropy to high accuracy at all temperatures.  With no additional parameters, the model also agrees with both experiment and molecular dynamics for the internal energy vs.~$T$ for Na.  With the calibrated transit model, V-T theory provides equations subject to \textit{ab initio} evaluation for thermodynamic properties of monatomic liquids.  This will allow the range of applicability of the theory, and its overall accuracy, to be determined.  More generally, the hypothesis of V-T theory, which divides the many-atom potential energy valleys into random and symmetric classes, can also be tested for its application beyond monatomic systems.

\end{abstract}


\maketitle

\section{Introduction}
\label{sec:Intro}

Our physical understanding of the motion of atoms in real liquids is notably deficient compared to 
lattice dynamics theory for real crystals.  The present work stands in a long line of research aimed 
at improving the liquid theory.  We shall consider only monatomic liquids, and equilibrium 
thermodynamic properties, though progress has recently been made in nonequilibrium theory as 
well \cite{DCW_PRE08a,WDC_ar05a}.

To rationalize experimental data, theory must be based on a physically realistic interatomic 
potential.  Pseudopotential perturbation theory provides such interatomic potentials for the nearly-free-electron metals \cite{H_66a}.  These potentials have been valuable in developing liquid dynamics 
theory, because the metals to which they apply are very well studied experimentally.  In the early 
days, pseudopotential perturbation theory was developed by comparing theory with experimental 
data for crystals.  Phonon dispersion curves were calculated for Na by Sham \cite{S_PRSLA65a}, and for Al 
by Harrison \cite{H_66a}.  Pseudopotential form factors were compared with Fermi surface data by 
Ashcroft  \cite{A_PL66a}, and compressibilities and binding energies were calculated by Ashcroft and 
Langreth \cite{AL_PR67a}.  Then, using a well-tested pseudopotential for Rb, Rahman demonstrated the 
ability of molecular dynamics (MD) to produce accurate results for statistical mechanical properties 
of the liquid \cite{R_PRL74a,R_PRA74a}.  A host of MD calculations followed, showing good agreement with experimental 
data for nearly-free-electron liquid metals.  These included the structure factor of alkali metals 
\cite{JKT_SSC76a,BTV_PRB93a}, thermodynamic properties of Na \cite{SSW_PR83a}, and structural and thermodynamic properties of 
alloys \cite{H_85a}.

The next step in liquid dynamics theory for real materials was the introduction of \textit{ab initio} MD 
by Car and Parrinello \cite{CP_PRL85a}.  This is still based on pseudopotentials in electronic structure theory, 
but the pseudopotential is treated numerically instead of as a perturbation.  The early development 
of computational methods is reviewed by Payne et al.~\cite{PTAAJ_RMP925a}.  Ultimately \textit{ab initio} MD 
calculations of the nuclei moving on the ground state adiabatic potential surface have produced 
results in good agreement with experiment for the melting properties of Al (de Wijs et al.~\cite{DKG_PRB98a}), structural and dynamic properties of liquid Fe under earth's core 
conditions (Alf\`{e} et al.~\cite{AKG_PRB00a}), and the pair distribution functions for groups IIIB--VIB elemental 
liquids (Kresse \cite{K_JNCS02a}; see also Chai et al.~\cite{CSHK_PRB03a} for Ge).  An extensive study of Na at high 
compression reveals significant electronic structure changes in the crystal (Neaton and Ashcroft 
\cite{NA_PRL01a}) and in the liquid (Raty et al.~\cite{RSB_N07a}), and a change from normal to anomalous melting 
(Gregoryanz et al.~\cite{GDSHM_PRL05a}).  Finally, \textit{ab initio} Monte Carlo calculation of liquid free 
energy is also being developed (Greeff and Liz\'{a}rraga \cite{GL_AIP07a}, Greeff \cite{G_JCP08a}).  

The goal of V-T theory is to develop a Hamiltonian formulation capable of analyzing the motion of 
atoms in the liquid state.  The theory is based on a fundamental hypothesis, and an advantageous 
decomposition of the atomic motion.  These propositions are formulated to rationalize a large body 
of experimental data for monatomic liquids \cite{W_PRE97b,CW_JPCM01a}.  The hypothesis, a symmetry classification of 
potential energy valleys, was argued from the rather universal value of the constant-volume 
entropy of melting for normal melting elements, $\Delta S \approx 0.8 k_B$/atom \cite{W_PRE97b,CW_JPCM01a}  (see also \cite{W_02a}, Sec.~22).  This 
universality implies the presence of a numerically dominant and uniform class of potential energy 
valleys, since a \emph{distribution} of potential energy properties could not be expected to produce 
the common entropy of melting \cite{W_PRE97b}.  The dominant uniform valleys are called random valleys.  
The atomic motion is decomposed into vibrations in the random valleys, plus complicated but less 
important transit motion, which carries the system between valleys.  This was argued from the 
nearly pure vibrational values of the atomic motion contribution to experimental specific heat and 
entropy \cite{W_PRE97b,CW_JPCM01a}.  These arguments are still valid, and still rationalize experimental data to an 
accuracy of a few percent.

In the original formulation, the transit partition function was set to $\mathcal{N}_r$, a constant 
representing the total number of random valleys \cite{W_PRE97b,CW_JPCM01a}.  This expresses the idea that all random 
valleys are equally accessible, and that every point in $3N$-dimensional configuration space 
belongs to one and only one potential energy valley \cite{W_PRE97b}.  However, this approximation for the 
transit partition function is not amenable to systematic improvement in order to develop a more 
accurate theory.  With the goal of improving transit theory, the available experimental high-temperature entropy data for elemental liquids were analyzed \cite{WCB_PRE09a}.  The result is a common curve 
of transit entropy for the liquids analyzed, which represents experiment to high accuracy \cite{WCB_PRE09a}.  
The purpose of this manuscript is to construct a statistical mechanics model for the transit partition 
function and to calibrate the model to the experimental transit entropy.  The model will then provide 
consistent calibrated equations for the transit free energy and all other transit thermodynamic 
functions.

In Sec.~\ref{sec:props}, the V-T propositions are described.  In Sec.~\ref{sec:Ham}, the V-T 
Hamiltonian is placed in the framework of the formally exact Hamiltonian of a condensed matter 
system (\cite{W_02a}, Sec.~4).  In Sec.~\ref{sec:partfunc}, the partition function is written, and equations for the 
internal energy and entropy are derived in classical statistical mechanics.  In Sec.~\ref{sec:mechsys}, the mechanical system representing the liquid is identified as the system whose 
potential energy surface consists of the random valleys.  The statistical mechanical properties of 
this system which must be accounted for in the transit partition function are described.  In Sec.~\ref{sec:partfuncmodel}, a model for the transit partition function is constructed, and is calibrated to 
the experimental transit entropy \cite{WCB_PRE09a}.  In Sec.~\ref{sec:entropy}, properties of the calibrated transit 
model are discussed, and in Sec.~\ref{sec:energy}, the model's ability to predict experimental 
internal energy data is shown for liquid Na.  The physical meaning of the two calibrated model 
parameters is also discussed.  Sec.~\ref{sec:predictions} discusses implications of two aspects of 
the present theory.  First, with the transit model developed here, liquid thermodynamic properties 
can be calculated without adjustable parameters to high accuracy.  Second, the symmetry 
classification of potential energy valleys is expected to be relevant beyond the liquid phase, and 
beyond monatomic systems.

\section{Formulation of V-T theory}
\label{sec:formulation}

\subsection{Basic propositions}
\label{sec:props}

In condensed matter theory, it is generally agreed that the potential energy surface is composed of 
intersecting many-atom potential energy valleys.  Our hypothesis divides these valleys into two 
classes, random and symmetric, with the following properties in the thermodynamic limit ($N 
\rightarrow \infty$) \cite{W_PRE97b,CW_JPCM01a}.
\begin{itemize}
\item[(i)] The random valleys are macroscopically uniform and numerically dominant.  Uniformity 
means for any macroscopic dynamical variable, the statistical mechanical average is the same for 
every random valley.  Numerically dominant means the liquid statistical mechanics at $T \geq T_m
$ 
is given entirely by the random valleys.
\item[(ii)] The symmetric valleys have a wide range of potential energy properties.  Qualitatively, 
potential properties of symmetric valleys range from crystal to liquid values \cite{WC_PRE99a,CW_PRE99a,DW_PRE07a,HBPLDCW_PRE09a}.  Symmetric 
valleys include those with microcrystalline structures; the single crystals are also included, and one 
of them has special status as the ground state structure. 
\end{itemize}

Current evidence supporting the symmetry classification will be summarized in Sec.~\ref{sec:predictions}.

Now, since the vibrational motion appears to dominate thermodynamic properties of monatomic liquids, we shall make this motion the leading term in our Hamiltonian.  For this purpose, we shall modify the vibrational motion to produce a tractable form.  Correction for this modification will then become part of the small but complicated transit Hamiltonian.

In a given random valley, at small displacements of the atoms from equilibrium, the system potential is quadratic in displacements.  We define the extended random valley as the extension to infinity of its harmonic potential surface.  The motion of atoms in an extended random valley is normal-mode vibrational motion, tractable in quantum and classical mechanics.  Because the extended random valleys are uniform in $N \rightarrow \infty$, a single such valley suffices for statistical mechanical calculations.  This property greatly simplifies the liquid dynamics theory.

So far we have followed the original formulation.  In preparation for a general treatment of transits, we shall make an accounting of the complete Hamiltonian for a condensed matter system (\cite{W_02a}, Sec.~4).

\subsection{Hamiltonian}
\label{sec:Ham}

The mechanical system has $N$ atoms in a volume $V$, with periodic boundary conditions on the 
atomic motion.  This motion is described by the Hamiltonian $\mathcal{H}$, where
\begin{equation} 
\mathcal{H} = \Phi_0^l + \mathcal{H}_{\rm vib} + \mathcal{H}_{\rm tr}.
\label{eq:Hdef}
\end{equation}
$\Phi_0^l(V)$ is the structural potential of random valleys in $N \rightarrow \infty$.  $
\mathcal{H}_{\rm vib}$ describes the vibrational motion in one (any) extended random valley,
\begin{equation}
\mathcal{H}_{\rm vib} = \sum_\lambda \left( \frac{p_\lambda^2}{2M} + \frac{1}{2}M\omega_
\lambda^2q_\lambda^2 \right).
\label{eq:Hvibdef}
\end{equation}
Here $M$ is the atomic mass, $q_\lambda$ and $p_\lambda$ are respectively the normal mode 
coordinates and momenta, and $\omega_\lambda$ are the normal mode frequencies, for $\lambda 
= 1, \ldots, 3N$.  Volume dependence of $\mathcal{H}_{\rm vib}$ is contained in the $\omega_
\lambda(V)$.  By definition of the structure, $\omega_\lambda^2 > 0$ for all $\lambda$, except for 
the three translational modes, for which $\omega_\lambda^2 = 0$ to numerical accuracy.

$\mathcal{H}_{\rm tr}$ is the transit Hamiltonian.  A transit occurs when the system crosses the 
boundary between two potential energy valleys.  The motion involves a small local group of atoms, 
and in equilibrium at $T \geq T_m$, transits are occurring at a high rate throughout the liquid.  The 
potential surface where the system moves in a transit, a \emph{transit surface}, differs locally from 
the extended random valley potential surface.  $\mathcal{H}_{\rm tr}$ is supposed to express this 
difference for all possible transit surfaces.  Our ultimate goal is to construct an explicit potential 
energy function for $\mathcal{H}_{\rm tr}$.  We shall not be able to do that here, but we shall be 
able to construct a simple statistical model that accounts for what is known about the transit 
contribution to thermodynamics.

In addition to the Hamiltonian contributions in Eq.~(\ref{eq:Hdef}), there are three terms which need not be considered here.  First is $\mathcal{H}_{\rm el}$, which expresses excitation of electrons from their ground state (\cite{W_02a}, Sec.~3).  The electronic ground state itself is the adiabatic potential for atomic motion, and is contained in the three terms on the right of Eq.~(\ref{eq:Hdef}).  The \emph{excitation} $\mathcal{H}_{\rm el}$ is determined by the electronic density of states evaluated for one (any) random structure \cite{footnote:dos}.  This term is important for liquid metals, contributing $1\%-10\%$ of the liquid internal energy and entropy.  This term is not explicitly included here, because we are discussing only the atomic-motion component of liquid dynamics.  However, experimental information presented here contains proper accounting of electronic excitations.

The next Hamiltonian contribution accounts for the \emph{interaction} between atomic motion and electronic excitation (\cite{W_02a}, Sec.~4).  From calculations for several metal crystals \cite{BCW_PRB05a,BWC_PRB06a}, we estimate for metallic liquids at $T \geq T_m$ that the adiabatic contribution dominates the nonadiabatic, and the adiabatic contribution to internal energy and entropy is on the order of the experimental error in these quantities.  We therefore neglect this term, because of its smallness.

Finally, in the actual random valley potential energy, there is anharmonicity \emph{not} associated with transits.  This is \emph{vibrational} anharmonicity.  In its contribution to thermodynamic functions, vibrational anharmonicity appears to be a much smaller effect than transits.  We therefore neglect vibrational anharmonicity, in the sense that we shall not attempt to model it explicitly.  Ultimately, however, our calibration of the transit partition function to experimental data will include vibrational anharmonicity.  The relative effect of this inclusion is presumed small \cite{footnote:vibanh}.

A note on the practical calibration of the Hamiltonian is useful.  The structural and vibrational parts of the Hamiltonian are calibrated with the parameters $\Phi_0^l(V)$ and $\{ \omega_\lambda(V) \}$, respectively.  These parameters can be calculated from model interatomic potentials \cite{DW_PRE07a,HBPLDCW_PRE09a} or from \textit{ab initio} electronic structure calculations \cite{BPCDWHL_08a}.  Since the calculations are done for finite systems, the results have finite-$N$ errors.  It is possible in principle to estimate finite-$N$ errors in the Hamiltonian calibration \cite{DW_PRE07a,HBPLDCW_PRE09a}.

\section{Statistical mechanics}
\label{sec:statmech}

\subsection{Partition function}
\label{sec:partfunc}

Let us write the partition function $Z$ corresponding to Eq.~(\ref{eq:Hdef}) for $\mathcal{H}$:
\begin{equation}
Z(V,T) = e^{-\beta \Phi_0^l(V)} Z_{\rm vib}(V,T) \, Z_{\rm tr}(V,T).
\label{eq:Zdef}
\end{equation}

$e^{-\beta \Phi_0^l(V)}$ is the structural partition function.  $Z_{\rm vib}(V,T)$ corresponds to $\mathcal{H}_{\rm vib}$, Eq.~(\ref{eq:Hvibdef}), and is fully quantum and possesses a classical limit.  $Z_{\rm tr}(V,T)$ corresponds to $\mathcal{H}_{\rm tr}$, and is the primary subject of this work.  Remaining factors in $Z(V,T)$, not written in Eq.~(\ref{eq:Zdef}), represent the three remaining Hamiltonian contributions mentioned in Sec.~\ref{sec:Ham}.

For simplicity, we shall treat the atomic motion by classical statistical mechanics.  This is quite accurate for most monatomic liquids at $T \geq T_m$, and the small quantum corrections can be estimated (\cite{W_02a}, Secs.~9 and 17).  Since the vibrational modes are orthogonal, $Z_{\rm vib}$ is a product of single normal-mode functions:
\begin{equation}
Z_{\rm vib}(V,T) = \prod_\lambda \sqrt{\frac{Mk_BT}{2\pi\hbar^2}} \int_{-\infty}^\infty \exp\left(-\frac{1}{2}\beta M \omega_\lambda^2 q_\lambda^2\right) dq_\lambda.
\label{eq:Zvibdef}
\end{equation}
The factors containing $\hbar$ express kinetic energy.  When the integrals are done, the result is
\begin{equation}
Z_{\rm vib}(V,T) = \prod_\lambda \left( \frac{k_B T}{\hbar \omega_\lambda} \right)  = \left[ \frac{T}{\theta_0^l(V)} \right]^{3N}.
\label{eq:Zvib}
\end{equation}
The second equality expresses the definition $\ln(k_B \theta_0^l) = \langle \ln(\hbar \omega_\lambda) \rangle$, where $\langle \, \cdots \rangle$ is the average over the set $\{\omega_\lambda\}$.

Complete expressions for the internal energy $U$ and the entropy $S$ are then
\begin{eqnarray}
U(V,T) & = & \Phi_0^l(V) + 3Nk_BT + U_{\rm tr}(V,T), \label{eq:U} \\
S(V,T) & =  & 3Nk_B\left\{\ln\left[T/\theta_0^l(V)\right]+1\right\} + S_{\rm tr}(V,T). \label{eq:S}
\end{eqnarray}
The primary theoretical quantities needed to evaluate these equations are $\Phi_0^l(V)$ for the energy and $\theta_0^l(V)$ for the entropy.  Additional moments of $\{\omega_\lambda\}$ are needed for the vibrational quantum corrections.

\subsection{Mechanical system of the liquid}
\label{sec:mechsys}

Let us define the random valley system as the mechanical system whose potential energy surface consists only of random valleys, and all of them.  In V-T theory, the random valleys make up the configuration space of the liquid.  The random valley system therefore represents the liquid, in the sense that statistical mechanics of the random valley system correctly describes the liquid state.  This property holds at all temperatures, at $T \geq T_m$ where the liquid is thermodynamically stable, and at $T < T_m$ where the liquid is metastable with respect to the crystal.  Statistical mechanics of metastable states is discussed in \cite{W_02a}, Sec.~27.

\begin{figure}
\includegraphics[width=\linewidth]{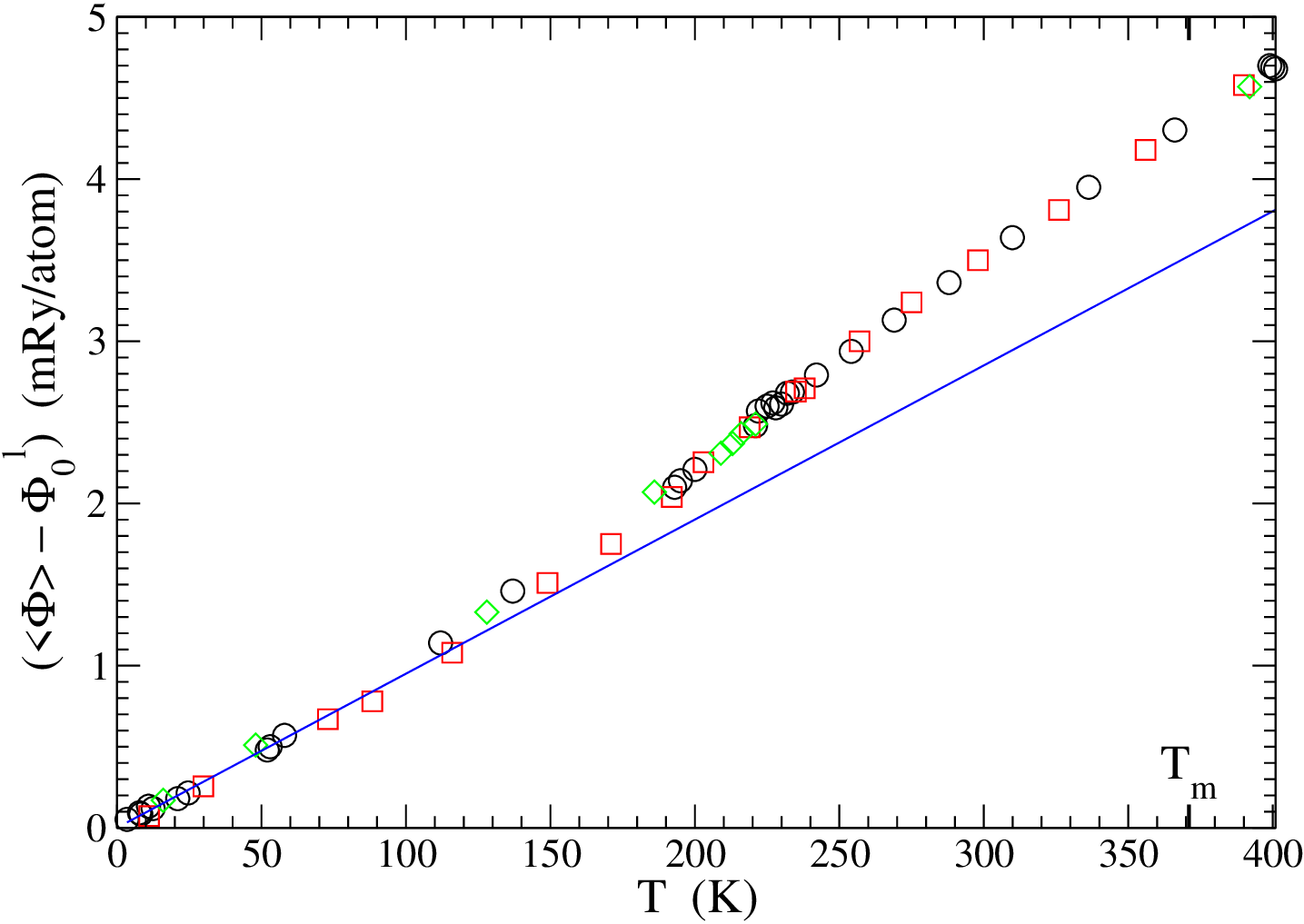}
\caption{(Color online) MD data for the mean potential energy $\langle\Phi\rangle - \Phi_0^l$ for the random valley system in Na as a function of temperature (symbols).  The volume is fixed at the volume of the liquid at melt at zero pressure.  The line is the mean vibrational potential energy of $(3/2)k_BT$ per atom; the difference is the energy due to transits.
  \label{fig:phivsT}
  }
\end{figure}

MD data for temperature dependence of the mean potential energy, $\langle\Phi\rangle - \Phi_0^l$, for the random valley system in Na is shown in Fig.~\ref{fig:phivsT}.  The points are from Fig.~4 of \cite{WC_PRE99a}.  The data are at the fixed volume $V^l_m$ of the liquid at melt at zero pressure, and the notation of volume dependence will be suppressed.  The zero-pressure melting temperature of Na is $371.0$ K.  

The mean vibrational potential energy is $(3/2)k_BT$ per atom, and is also graphed in Fig.~\ref{fig:phivsT}.  The transit internal energy per particle, $U_{\rm tr}$, is the difference of the two curves in Fig.~\ref{fig:phivsT}:
\begin{equation}
U_{\rm tr} = \left( \langle\Phi\rangle - \Phi_0^l \right) - \frac{3}{2}Nk_BT.
\label{eq:Utrfromdata}
\end{equation}
Notice this is the potential energy component of Eq.~(\ref{eq:U}).  We shall not attempt to use this information to calibrate a model of $Z_{\rm tr}$ for Na.  Rather, we shall use the information to characterize a statistical mechanical model for monatomic liquids in general.

The following significant properties of the random valley system were observed in \cite{WC_PRE99a}, and are reflected in Fig.~\ref{fig:phivsT}.
\begin{enumerate}
\item[(a)] At low temperatures, the equilibrium MD system is observed to remain in a single random valley for a very long time.  The mean potential energy corresponds to vibrational motion, with anharmonicity too small to measure.  Equation (\ref{eq:Utrfromdata}) implies $U_{\rm tr}=0$.
\item[(b)] Upon warming the MD system, a narrow temperature range is reached where $U_{\rm tr}$ is observed to increase from zero.  In the same temperature range, the self diffusion coefficient $D$ increases from zero. 
Since self diffusion results entirely from transits, the coincident appearance of nonzero $D$ and $U_{\rm tr}$ confirms $U_{\rm tr}$ as due to transits.
\item[(c)] The high-temperature states in Fig.~\ref{fig:phivsT} are at $T  > T_m$.  The MD values of $\langle\Phi\rangle$ for these states agree with experimental data for liquid Na.
\end{enumerate}

Let us interpret the preceding observations in terms of the statistical mechanics of the random valley system, and include the transit entropy in the discussion.  This will provide the list of transit properties we must account for in a statistical mechanics model for $Z_{\rm tr}$.

At low temperatures, the random valley system becomes trapped in a single random valley, where the motion is entirely vibrational.  Since the transit surfaces are not visited by the system, both $U_{\rm tr}$ and $S_{\rm tr}$ are zero.  Upon warming, the transit rate increases, and at some (qualitative) temperature, both $U_{\rm tr}$ and $S_{\rm tr}$ increase from zero.  This process is thermally activated.  The process should not be thought of as melting, nor its inverse as freezing.  Since the random valleys all have the same potential energy parameters, no first-order phase transition is present.  With a further increase of temperature, $U_{\rm tr}$ and $S_{\rm tr}$ saturate and then begin to decrease.  This behavior is not shown in Fig.~\ref{fig:phivsT}, but is present in the experimental internal energy and entropy data of monatomic liquids (Sec.~\ref{sec:caltransitmodel}).

\subsection{Model for the transit partition function}
\label{sec:partfuncmodel}

The Hamiltonian parameters are $V$-dependent, while $T$ dependence is contained in $\beta=(k_BT)^{-1}$.  Volume dependence will remain suppressed.  Each normal-mode configuration integral, $Q_\lambda$, has vibrational contribution given by the integral in Eq.~(\ref{eq:Zvibdef}).  To include transits, a separate transit-surface segment is added to the integral, so that
\begin{eqnarray}
Q_\lambda(T) & = & \int_{-\infty}^\infty \exp\left(-\frac{1}{2}\beta M \omega_\lambda^2 q_\lambda^2\right) dq_\lambda \nonumber \\ 
& & + \left( \int_{-c_\lambda}^{-b_\lambda} + \int_{b_\lambda}^{c_\lambda} \right) \exp\left(-\beta \epsilon_\lambda \right) dq_\lambda.
\label{eq:Qldef}
\end{eqnarray}
The transit surfaces are $b_\lambda \leq |q_\lambda| \leq c_\lambda$, where $0 < b_\lambda < c_\lambda$, and the transit surfaces have potential energy $\epsilon_\lambda > 0$.  Evaluation of Eq.~(\ref{eq:Qldef}) gives
\begin{equation}
Q_\lambda(T)=\sqrt{\frac{2\pi k_BT}{M\omega_\lambda^2}} \left[ 1 + \sqrt{\frac{2M\omega_\lambda^2 d_\lambda^2}{\pi k_BT}}\exp(-\beta\epsilon_\lambda)\right],
\label{eq:Ql}
\end{equation}
where $d_\lambda=c_\lambda-b_\lambda$.  The factor outside the brackets is the vibrational contribution, which must be taken with the kinetic energy contribution to produce $Z_{\rm vib}$, Eq.~(\ref{eq:Zvib}).  Then the bracket in Eq.~(\ref{eq:Ql}) is the single normal-mode contribution to $Z_{\rm tr}$, so the total $Z_{\rm tr}$ is then
\begin{equation}
Z_{\rm tr}(T)=\prod_\lambda \left[ 1 + \sqrt{\frac{2M\omega_\lambda^2 d_\lambda^2}{\pi k_BT}}\exp(-\beta\epsilon_\lambda)\right].
\label{eq:Ztr}
\end{equation}
We shall next simplify the model to a form appropriate to the work at hand.

In Eq.~(\ref{eq:Ztr}), each normal mode has two energy parameters, namely $\epsilon_\lambda$ and $M \omega_\lambda^2 d_\lambda^2$.  We shall set each of these the same for every mode.  It is convenient to use only a single energy parameter $\epsilon$, as follows:
\begin{eqnarray}
\epsilon_\lambda & = & \epsilon, \nonumber \\
\sqrt{\frac{2}{\pi} M \omega_\lambda^2 d_\lambda^2} & = & \mu \sqrt{\epsilon}, 
\label{eq:epscale} 
\end{eqnarray}
where $\mu > 0$ is a dimensionless parameter.  Equation (\ref{eq:Ztr}) becomes
\begin{eqnarray}
Z_{\rm tr}(T) & = & \left[ 1 + h(T) \right]^{3N}, \label{eq:simpleZtr} \\
h(T) & = & \mu \sqrt{\beta\epsilon} \exp(-\beta\epsilon).
\label{eq:hTdef}
\end{eqnarray}
The $T$ dependence of $Z_{\rm tr}(T)$ is entirely contained in $\beta\epsilon$.  This is the source of temperature scaling in the transit properties.

The thermodynamic functions now follow.  The Helmholtz free energy is $F_{\rm tr}$, the internal energy and entropy are respectively $U_{\rm tr}$ and $S_{\rm tr}$, and the constant-volume specific heat is $C_{\rm tr}$.  These are given by
\begin{eqnarray}
F_{\rm tr} & = & -3Nk_BT\ln\left[1+h(T)\right] \label{eq:Ftr} \\
U_{\rm tr} & = & 3Nk_BT\frac{\left(\beta\epsilon - \frac{1}{2}\right)h(T)}{1+h(T)} \label{eq:Utr} \\
S_{\rm tr} & = & 3Nk_B\left[ \ln\left[1+h(T)\right] + \frac{\left(\beta\epsilon - \frac{1}{2}\right)h(T)}{1+h(T)}\right] \label{eq:Str} \\
C_{\rm tr} & = & 3Nk_Bh(T)\left[\frac{(\beta\epsilon - \frac{1}{2})^2}{(1+h(T))^2} - \frac{\frac{1}{2}}{1+h(T)}\right]. \label{eq:Ctr}
\end{eqnarray}

Analysis of the available high-temperature experimental entropy data for monatomic liquids has yielded a common curve for $S_{\rm tr}$ as a function of $T/\theta_{\rm tr}$, where $\theta_{\rm tr}$ is a scaling temperature for each liquid \cite{WCB_PRE09a}.  The common curve has a maximum value of $S_{\rm tr}=0.8Nk_B$, located at $T=\theta_{\rm tr}$.  Since $\theta_{\rm tr}$ is a material parameter, we must calibrate for each liquid independently.  Since the $T$ dependence of $S_{\rm tr}(T)$ is contained in $\beta\epsilon$, according to Eqs.~(\ref{eq:hTdef}) and (\ref{eq:Str}), the model $S_{\rm tr}(T)$ will scale with $T/\theta_{\rm tr}$ if we set
\begin{equation}
\epsilon = \nu k_B \theta_{\rm tr},
\label{eq:nudef}
\end{equation}
where $\nu$ is a dimensionless parameter.  Calibration will require the determination of $\mu$ and $\nu$ for each liquid.  Denote by $\chi$ the maximum value of $S_{\rm tr}(T)$, which occurs at $T=\theta_{\rm tr}$, or $\beta\epsilon=\nu$.  Then the value and slope of $S_{\rm tr}(\theta_{\rm tr})$ are calibrated to
\begin{eqnarray}
S_{\rm tr}(\theta_{\rm tr}) & = & S_{\rm tr}(\mu,\nu) = \chi, \nonumber \\
C_{\rm tr}(\theta_{\rm tr}) & = & C_{\rm tr}(\mu,\nu) = 0.
\end{eqnarray}
So $\mu$ and $\nu$ are functions of $\chi$ only, independent of $\theta_{\rm tr}$.  If $\chi$ is the same for all liquids, then $\mu$ and $\nu$ are the same as well.  We set $\chi=0.8Nk_B$ for monatomic liquids, and find
\begin{eqnarray}
\mu & = & 0.53221, \nonumber \\
\nu & = & 1.26452.
\label{eq:setmunu}
\end{eqnarray}

Ultimately, the statistical mechanics model is quite simple.  The complete potential energy surface is modeled by two components: vibrational and transit.  That the vibrational surface lies \emph{below} the transit surface makes $U_{\rm tr}$ and $S_{\rm tr}$ go to zero at low temperatures [$\epsilon > 0$ in Eqs.~(\ref{eq:hTdef})-(\ref{eq:Str})].  That the vibrational surface continues \emph{above} the transit surface causes $U_{\rm tr}$ and $S_{\rm tr}$ to saturate and then decrease with increasing temperature [the transit term in Eq.~(\ref{eq:Ql})].  In this way the model possesses the required characteristics, as listed in the last paragraph of Sec.~\ref{sec:mechsys}.

\section{Discussion of the calibrated transit model}
\label{sec:caltransitmodel}

\subsection{Transit entropy}
\label{sec:entropy}

\begin{figure}
\includegraphics[width=\linewidth]{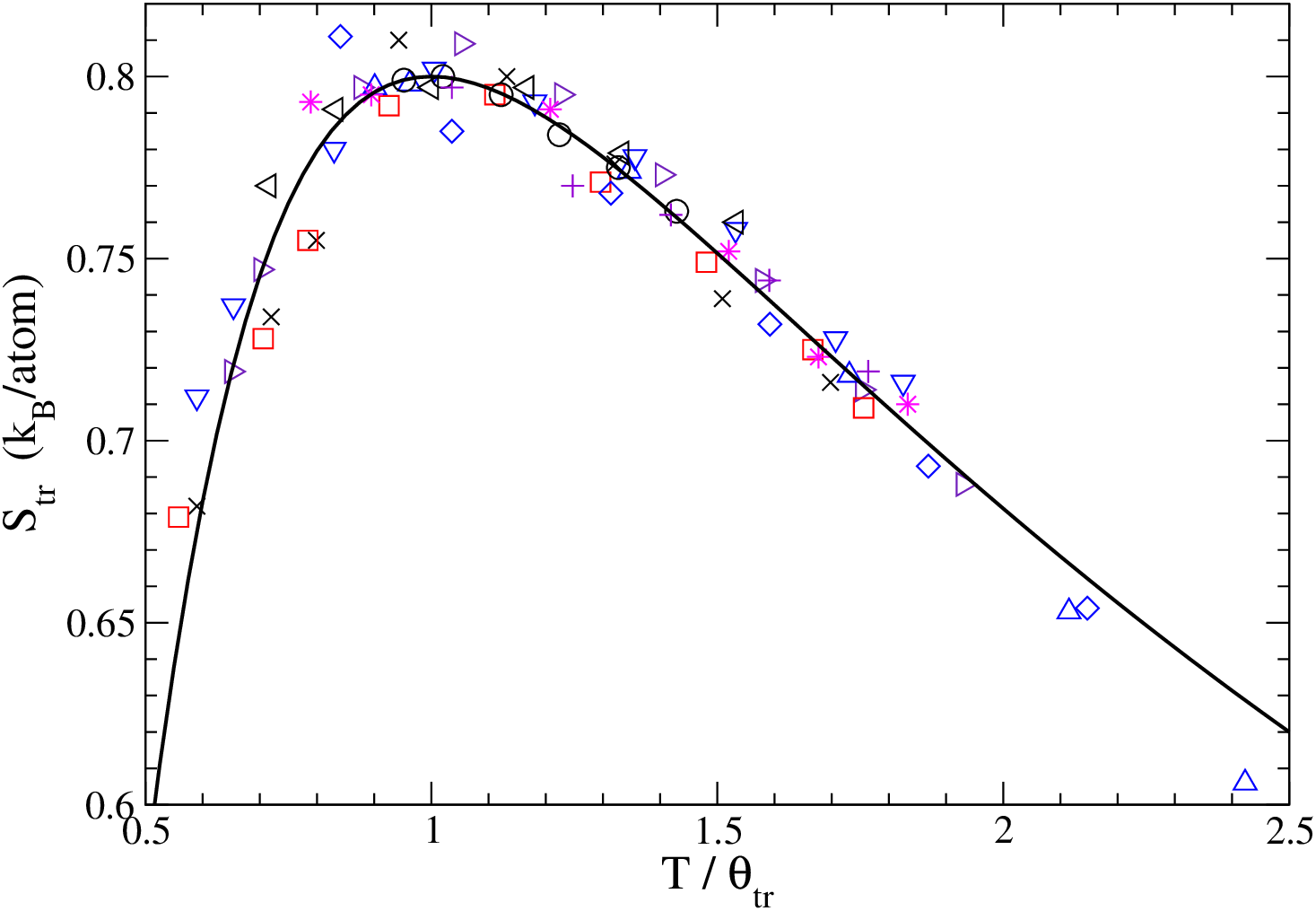}
\caption{(Color online) Experimental transit entropy data for ten elemental liquids as a function of temperature scaled by the characteristic transit temperature $\theta_{\rm tr}$.  The data are from Fig.~1 of \cite{WCB_PRE09a}.  The solid line is the transit model presented here, Eq.~(\ref{eq:Str}), calibrated as indicated in Eq.~(\ref{eq:setmunu}).
  \label{fig:Sdatwmodel}
  }
\end{figure}

Figure \ref{fig:Sdatwmodel} shows the experimental transit entropy data for ten elemental liquids at the fixed volume $V^l_m$ (the volume of the liquid at melt at zero pressure), as a function of $T/\theta_{\rm tr}$.  Also shown is $S_{\rm tr}(T/\theta_{\rm tr})$ from the statistical mechanics model, Eq.~(\ref{eq:Str}), with the calibration Eq.~(\ref{eq:setmunu}).  The agreement of model with experiment is excellent; scatter of the points from the line is of order $0.01\, k_B$/atom, or around $0.1\%$ of the total entropy.  However, we do not suppose the model is entirely ``correct'' to this level of accuracy.  Systematic errors of the experimental entropy data are larger than the scatter in Fig.~\ref{fig:Sdatwmodel}, and these errors can affect the shape of the $S_{\rm tr}(T)$ curve.  This uncertainty is reflected in the estimate of up to $10\%$ error in the fitted values of $\theta_{\rm tr}$ \cite{WCB_PRE09a}.

To understand Fig.~\ref{fig:Sdatwmodel}, one needs to know how it is made \cite{WCB_PRE09a}.  The zero-pressure experimental data $S_{\rm expt}(V,T)$ are corrected, by means of additional experimental data, to the fixed volume $V^l_m$.  This produces a data set for  $S_{\rm expt}(V^l_m,T)$ for each liquid.  This data set is then fitted to the right side of Eq.~(\ref{eq:S}), with $\theta_0^l$ a variable parameter.  In the fitting process, the maximum of each $S_{\rm tr}(V^l_m,T)$ curve is fixed at $0.8\, k_B$/atom, and the temperature at the maximum is denoted $\theta_{\rm tr}$.  The fitted characteristic temperatures $\theta_0^l$ and $\theta_{\rm tr}$ are listed in \cite{WCB_PRE09a}.

The presence of temperature scaling in the statistical mechanics model is informative.  It results from the simplicity of the model, in that $T$ appears in only one form in $Z_{\rm tr}$, namely in the form $\beta\epsilon$ [see Eqs.~(\ref{eq:simpleZtr}) and (\ref{eq:hTdef})].  The previously-discovered scaling temperature $\theta_{\rm tr}$ is then introduced via Eq.~(\ref{eq:nudef}), leaving $\nu$ as the parameter in place of $\epsilon$.  The following dimensionless functions are then functions only of $T/\theta_{\rm tr}$: $\beta\epsilon$, $h$, $S_{\rm tr}/k_B$, $C_{\rm tr}/k_B$, $F_{\rm tr}/k_BT$, and $U_{\rm tr}/k_BT$.  This scaling behavior should prove useful in studying the transit thermodynamic functions, especially since they are relatively small contributions.

Finally in Fig.~\ref{fig:Sdatwmodel}, we note that the model rises slightly above the data at $T \gtrsim 2\theta_{\rm tr}$.  This discrepancy can be due to experimental errors in the data analysis.  However, the discrepancy might be significant, and due to transits at very high potential energy.  Here, the major transit correction is the removal of the vibrational surface that was extended beyond the random valley boundary.  That correction is modeled in a study of liquid Hg at high temperatures \cite{W_PRE98b}.  This effect would break the $T/\theta_{\rm tr}$ scaling of $S_{\rm tr}$.  But the effect seen in Fig.~\ref{fig:Sdatwmodel} is too small to justify modeling.

\subsection{Transit internal energy}
\label{sec:energy}

\begin{figure}
\includegraphics[width=\linewidth]{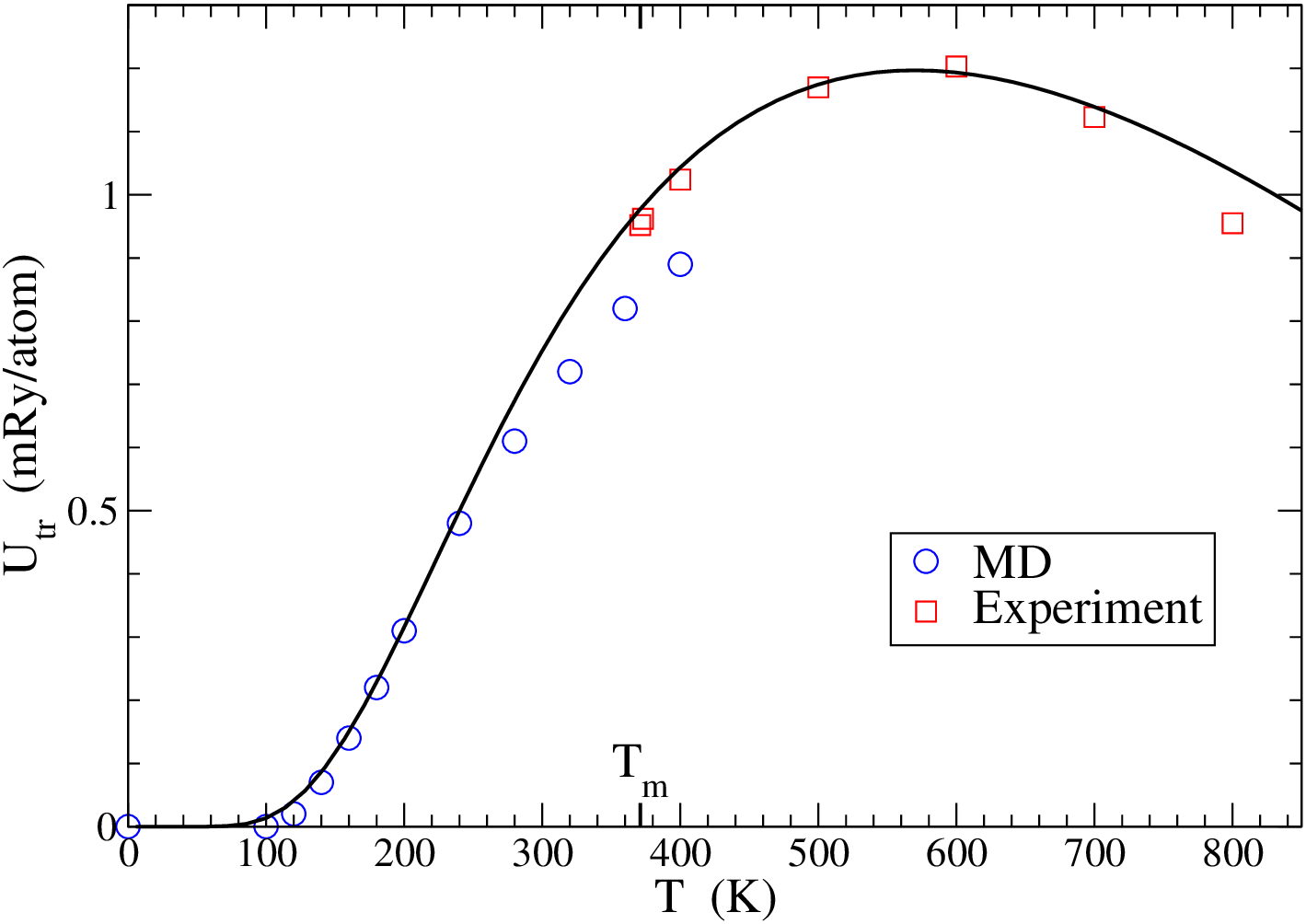}
\caption{(Color online) Transit energy per atom for Na, determined from MD calculations and experiment, as a function of temperature at fixed volume.  The solid line is the transit model presented here, Eq.~(\ref{eq:Utr}), calibrated as indicated in Eq.~(\ref{eq:setmunu}).
  \label{fig:Udatwmodel}
  }
\end{figure}

Experimental data for the zero-pressure internal energy $U(V,T)$ for liquid Na are corrected to the fixed volume $V^l_m$, and the experimental $U_{\rm tr}(V^l_m,T)$ is extracted by means of Eq.~(\ref{eq:U}).  For this, the value $\Phi_0^l(V^l_m)=0.33 \pm 0.05$ mRy/atom is estimated from early MD data \cite{SSHW_PR82a,SW_PRB84a}.  $\Phi_0^l$ is measured relative to the thermodynamic zero of energy, which is the energy of the crystal at zero temperature and pressure.  The experimental data for $U_{\rm tr}(V^l_m,T)$ are graphed in Fig.~\ref{fig:Udatwmodel}.  MD data for $U_{\rm tr}(V^l_m,T)$ are obtained from Fig.~\ref{fig:phivsT}, by means of Eq.~(\ref{eq:Utrfromdata}), and are also graphed in Fig.~\ref{fig:Udatwmodel}.  The model curve is calculated from Eqs.~(\ref{eq:Utr}) and (\ref{eq:setmunu}), together with $\theta_{\rm tr} = 570$ K for liquid Na \cite{WCB_PRE09a}.

In Fig.~\ref{fig:Udatwmodel}, the model curve rises above experiment at $T \gtrsim 800$ K.  This is due to error in the volume correction of the experimental energy.  That volume correction is extremely difficult to evaluate, hence becomes inaccurate at a relatively low temperature.  Also in Fig.~\ref{fig:Udatwmodel}, at $T_m$, the MD curve is lower than experiment by $1.4\%$ of the total experimental internal energy.  This is likely due to error in the Na interatomic potential, plus a few smaller contributing errors.  Overall, the discrepancies in Fig.~\ref{fig:Udatwmodel} are remarkably small, and can be ignored in our analysis.

Let us first compare the model with experiment in Fig.~\ref{fig:Udatwmodel}.  Since the model agrees with experiment for $S_{\rm tr}(T)$ at $T \geq T_m$, Fig.~\ref{fig:Sdatwmodel}, it must also agree with experiment for $U_{\rm tr}(T)$ at $T \geq T_m$, with the possible exception of an error in a constant of integration.  The constant of integration can be taken as the value of $U_{\rm tr}(T_m)$.  This being given correctly by the model constitutes a verification of the model, independent of the verification provided by the entropy in Fig.~\ref{fig:Sdatwmodel}.

Let us next compare the model with MD in Fig.~\ref{fig:Udatwmodel}.  The comparison is proper since both are classical.  One immediately sees that $U_{\rm tr}(T)$ increases from zero at the same temperature for both the model and MD.  Notice also that $S_{\rm tr}(T)$ will increase from zero at the same temperature as $U_{\rm tr}(T)$ does, both for MD data and for the model.  In Fig.~\ref{fig:Udatwmodel}, agreement of the model with MD at $T < T_m$ is not entirely independent of the agreement with experiment at $T \geq T_m$.  It is nevertheless quite satisfactory that the transit model, calibrated at $T \geq T_m$, goes to zero at the correct temperature (for Na) well below $T_m$.

In Sec.~\ref{sec:mechsys}, it was argued that $S_{\rm tr}(T)$ and $U_{\rm tr}(T)$ must vanish at temperatures where transits are not thermally activated in the random valley system.  Hence, $S_{\rm tr}(T)$ contains no additive constant.  This implies that the random valley multiplicity does not contribute to $S_{\rm tr}(T)$, so that the effective number $\mathcal{N}_r$ of random valleys satisfies $\ln \mathcal{N}_r < \mathcal{O}(N)$.  The agreement between theory and experiment in Figs.~\ref{fig:Sdatwmodel} and \ref{fig:Udatwmodel} provides support for this conclusion.

The model parameters have physical meaning, which is sharpened by the calibration process.  Our tentative presumption is that this discussion applies to monatomic liquids in general.  The potential energy of each effective transit surface is $\epsilon$, measured from the structural potential $\Phi_0^l$, the same level from which vibrational energy is measured.  From the calibration of $\nu$, Eq.~(\ref{eq:setmunu}), we have $\epsilon = 1.265 k_B\theta_{\rm tr}$.  Hence the transit surface is easily accessible at liquid temperatures.  $\mu$ is related to the effective length of the transit surface in each $q_\lambda$ direction, and $\mu$ fixes the magnitude of $h(T)$, Eq.~(\ref{eq:hTdef}).  The ratio of transit to vibrational contributions in the internal energy and entropy is of order $h(T)$, which is small compared to $1$ at all $T$.  The ratio can be calculated exactly from the equations of Sec.~\ref{sec:statmech}.  This provides a quantitative statement of our introductory message, that the transit contribution to liquid thermodynamic properties is small, but important for accurate work.

\section{Theoretical predictions and verifications}
\label{sec:predictions}

With the calibrated transit model of Sec.~\ref{sec:statmech}, it is possible to evaluate thermodynamic properties of elemental liquids without adjustable parameters.  We consider specifically the internal energy and entropy, whose V-T theory formulas are given in Eqs.~(\ref{eq:U}) and (\ref{eq:S}).  To date, three independent tests verify these formulas to an accuracy within experimental error, and are therefore consistent with zero theoretical error.  By comparing \textit{ab initio} calculations of $\theta_0^l$ with the values determined by fitting the experimental entropy, it is shown that Eq.~(\ref{eq:S}) agrees with experiment for Na and Cu \cite{WCB_PRE09a}.  Equation (\ref{eq:U}), with MD evaluation of $\Phi_0^l$ and with $U_{\rm tr}(T)$ from Eq.~(\ref{eq:Utr}), agrees with experiment for the internal energy of Na (Fig.~\ref{fig:Udatwmodel}).  Continued testing in this way will reveal the overall accuracy of the theory, and will uncover any cases where the theory needs significant correction.

In comparing theory and experiment for the liquid entropy, the temperature dependence of experimental entropy is already accounted for by the transit model for the ten liquids in Fig.~\ref{fig:Sdatwmodel}.  For other liquids, temperature dependence of experimental entropy is independent information, so that the entire temperature dependence of the internal energy or entropy will test the theory.  In the analysis of experimental data reported in this paper, electronic excitation contributions are evaluated from free electron theory.  A more accurate calculation is based on the electronic density of states evaluated for a random structure (Sec.~\ref{sec:Ham}).  The random structure density of states will be \emph{necessary} for transition metals, where electronic excitation contributions are much larger than in the nearly-free-electron metals.  This provides an additional theoretical prediction which can be tested.  The same theory was used to isolate the anharmonic vibrational contribution to entropy in the transition metal crystals \cite{EWW_PRB92a}.

Our development of monatomic liquid dynamics theory is strongly based on the symmetry classification of potential energy valleys, Sec.~\ref{sec:props}.  The same random valleys important for liquid theory will continue to dominate at higher temperatures, where the system undergoes the broad liquid-to-gas transition \cite{footnote:liq-to-gas}.  Moreover, all the potential valleys, random and symmetric, will contribute to the description of amorphous solids at temperatures well below $T_m$.  In view of this extended application of the symmetry classification hypothesis, it is worthwhile to investigate to what classes of materials it applies.

The symmetry classification hypothesis is now well verified for Na at the density of the liquid at melt \cite{DW_PRE07a,HBPLDCW_PRE09a}.  The complete distribution of structural potentials $\Phi_0$ shows the narrow but dominant random peak, the broad symmetric distribution, and the crystal \cite{HBPLDCW_PRE09a}.  Other measures can distinguish randoms from symmetrics, e.g., Voronoi analysis and pair correlations \cite{CW_PRE99a}.  Work in progress shows the dominance and uniformity of random structures for Al and Cu.  The logical next step is to test the hypothesis for monatomic systems in general.  The characteristic random structures are expected to vary with volume for one element, and to vary from one element to another.

In principle, the symmetry classification of potential energy valleys should also apply to more complicated systems, e.g., to alloys, compounds, and molecular systems.  Evidence has been cited for the presence of a dominant and uniform class of potential valleys, i.e., the random class, in a variety of MD systems (see \cite{DW_PRE07a} and references [56-62] quoted there).  If the random valley class is found to be present in complex liquids, we shall have moved a step closer to a Hamiltonian formulation for such systems.

\section{Acknowledgments}

We appreciate helpful discussions with B.~Clements, C.~Greeff, and T.~Peery.  This work was funded by the U.S.~Department of Energy under Contract No.~DE-AC52-06NA25396.

\bibliography{VTref_17Dec2009}

\end{document}